# DFT+DMFT study on pressure-induced valence instability of CeCoSi

Shuai-Kang Zhang,[1, *] Yuanji Xu,[2] Guojun Li,[1, †] Junshuai Wang,[1] Zhongpo Zhou,[1] and Yipeng An[1, ‡]

[1] *School of Physics, Henan Normal University, Xinxiang 453007, China*

[2] *Institute for Applied Physics, University of Science and Technology Beijing, Beijing 100083, China*

[*]zhangshuaikang@htu.edu.cn

[†]liguojun@htu.edu.cn

[‡]ypan@htu.edu.cn

**Abstract:** Rare-earth compounds $R$CoSi exhibit unique properties, with distinct structural behaviors depending on whether $R$ is a light, middle or heavy rare-earth element. Among them, CeCoSi undergoes a structural phase transition under high pressure, with the phase transition pressure increasing as temperature rises. Some experimental studies suggest that the transition is closely related to the behavior of Ce-4$f$ electrons. In this work, we systematically studied the evolution of the electronic structure of CeCoSi with temperature and pressure. First, we used the density functional theory (DFT) + dynamical mean field theory (DMFT) to calculate the energy-volume curve of CeCoSi, which was in good agreement with the experimental results and far superior to the DFT method. Next, we studied the electronic structure of CeCoSi under different pressures and temperatures using DFT+DMFT. Our results show that CeCoSi is a Kondo metal with hybridization of Ce-4$f$ and Co-3$d$. As pressure increases, the renormalization factor $Z$ of Ce-4$f_{5/2}$ increases, the occupancy number of Ce-4$f$ electrons decreases, and CeCoSi transitions to a mixed-valence state at ~5.5 GPa in 100 K. The pressure of the quantum phase transition $P_Q$ is slightly higher than the experimentally observed structural phase transition pressure $P_S$, and the $P_Q$ increases with increasing temperature, which is consistent with the behavior of $P_S$ in experiment. In addition, the hybridization strength of Ce-4$f$ in the mixed-valence state is significantly greater than in the Kondo metal state. Our results suggest that the valence instability of Ce-4$f$ is the cause of the structural phase transition. As pressure increases, Ce-4$f$ electrons delocalize and CeCoSi transitions to mixed-valence state. This valence instability may cause redistribution of electron density, thus inducing a structural phase transition. Our work reveals the cause of the structural phase transition of CeCoSi under high pressure.

## I. INTRODUCTION

Rare-earth compounds, due to the interaction of localized $f$ electrons, exhibit complex phenomena such as quantum criticality, unconventional superconductivity, and heavy fermions, making them a central focus of condensed matter physics[1-7]. And, external factors like pressure, temperature, and magnetic field can induce novel physical behaviors, providing a unique platform for the study of correlated electron interactions[8-12].

*RTMX* (*R* = rare-earth metal, *TM* = transition metal, *X* = *p*-element) compounds exhibit different crystal structures depending on their composition of element. In *R*CoSi compounds, light or middle rare-earth elements (*R* = La, Ce, Pr, Nd, Sm, Gd, Tb) form a tetragonal CeFeSi-type structure (space group No. 129, *P*4/*nmm*)[13-15]; heavy rare-earth elements (*R* = Dy, Ho, Er, Tm, Lu) form a orthorhombic TiNiSi-type structure (space group No. 62, *Pnma*)[16,17]. These structural variations are likely due to the different ionic radii of the *R* ions. Hence, studying pressure-induced structural phase transitions in *R*CoSi compounds with light rare-earth elements may provide insight into the structural differences across the *R*CoSi family.

Among them, the light rare-earth *R*CoSi compound CeCoSi exhibits some unique properties under pressure. CeCoSi adopts a tetragonal CeFeSi-type crystal structure and lacks local inversion symmetry at the Ce site, as shown in Fig. 1(a). It exhibits a distinctive low-temperature pressure phase diagram, with a paramagnetic state above $T_0 \sim 12$ K, long-range order below $T_0$, and antiferromagnetic order at $T_N \sim 10$ K[18-21]. As pressure increases, $T_N$ gradually decreases and disappears around 1.4 GPa, while $T_0$ rises with pressure, sharply increasing to 40 K at $P \sim 1.5$ GPa and decreases and disappears at $P \sim 2.15$ GPa. Some studies suggest that the magnetic ordering near $T_N \sim T_0$ may stem from antiferroquadrupolar order[22]. Additionally, CeCoSi displays another unique propertie. Y. Kawamura *et al.* conducted X-ray diffraction and resistivity measurements on CeCoSi up to 8 GPa and observed a pressure-induced structural phase transition[23]. At 300 K, the phase transition pressure $P_s$ is approximately 4.9 GPa, decreasing to 3.6 ± 0.4 GPa at 10 K. These transitions are not related to low-temperature magnetic ordering. Studies of LaCoSi and PrCoSi under pressure, however, revealed no structural phase transitions up to 9 GPa[24], suggesting that the phase transition in CeCoSi near 5 GPa is linked to the behavior of Ce-4$f$ electrons.

To investigate the origins of the pressure-induced structural phase transition in CeCoSi, we employed DFT and DFT+DMFT methods to study the Ce-4$f$ occupation, spectral function, renormalization factor $Z$, Fermi surface, and hybridization function with different pressures and temperatures. Since there is still a lack of experimental research on the structure of high-pressure phase, we used the low-pressure tetragonal structure in calculations. Our results show that CeCoSi is a Kondo metal at ambient pressure and exhibits significant valence instability, which is highly sensitive to pressure. As pressure increases, CeCoSi undergoes a quantum phase transition from Kondo metal to mixed-valence state. At 100 K, this transition occurs around 5.5 GPa, with the quantum phase transition pressure $P_Q$ increasing with temperature, reaching about 6.5 GPa at 300 K. The $P_Q$ is slightly higher than the experimentally observed $P_s$, with both exhibiting same trends with temperature[23]. CeCoSi undergoes a structural phase transition before the quantum phase transition, indicating that the structural phase transition is driven by the valence instability of Ce-4$f$ electrons. Furthermore, we calculated the hybridization function at different pressures and temperatures, finding that, the hybridization strength of the Kondo metal state is significantly smaller than that of mixed-valence state. This indicates that with increasing pressure, the Ce-Co distance decreases, resulting in enhanced hybridization and delocalization of Ce-4$f$, and CeCoSi transforms into a mixed-valence state. The instability of the valence state likely causes a redistribution of electron density, thereby inducing the structural phase transition. Our work reveals the cause of the structural phase transition of CeCoSi under high pressure.

## II. COMPUTATIONAL METHODS

DFT Calculations were performed using the full-potential linearized augmented plane-wave (FLAPW) method implemented in the WIEN2k code[25]. The muffin-tin radii of Ce, Co, and Si atoms were set to 2.50 a.u., 2.13 a.u., and 1.89 a.u., respectively, at all pressures. The muffin-tin radii multiply the plane-wave cutoff $R_{mt}K_{max}$ and the $k$-point number in the Brillouin zone was set to 8.0 and 4000 for all calculations, at all pressures. The exchange-correlation functional employed was the Perdew–Burke–Ernzerhof (PBE) functional[26], and the spin-orbit coupling (SOC) effect was considered in a second-order variational manner in WIEN2k code[25].

The DFT+DMFT method combines the realistic band-structure calculations in DFT with the nonperturbative manybody treatment of local interaction effects in DMFT[27,28]. This method has been successfully applied to several strongly correlated materials. The DFT and DMFT parts of the calculation are solved separately using the WIEN2k and eDMFT code[29-31]. The same parameters for $R_{mt}K_{max}$, $k$-grids, and exchange-correlation functional were used in the DFT+DMFT calculations as in the DFT calculations. the onsite interaction parameters Hubbard $U$ and Hund exchange $J$ for Ce-4$f$ were set to $U = 6.0$ eV and $J = 0.7$ eV, which are commonly used values[32-34]. Given the relatively large bandwidth of Co-3$d$ electrons, the strong correlation effects were only considered for the Ce-4$f$ orbitals. The SOC effect was included in the same manner as in the DFT calculations. To avoid the negative sign problem, the off-diagonal components of the Ce-4$f$ hybridization function were neglected, as they do not significantly affect the results. The multi-orbital Anderson impurity model was solved using the continuous-time quantum Monte Carlo (CT-HYB) impurity solver[35]. The energy window for Ce-4$f$ orbital projection was set to -10 ~ 10 eV relative to Fermi energy ($E_F$). To reduce computational costs, the local Hilbert space was truncated severely, considering only the states $N \in [0, 1, 2, 3]$. The self-energy in real frequency was obtained through analytic continuation using the maximum entropy method[36].

The lattice constants of CeCoSi at pressures ranging from 0 to 4 GPa used in the calculations were taken from experimental data at 300 K. The lattice constants for other pressures were estimated by linear fitting of the experimental data[23].

### III. RESULTS AND DISCUSSION

To verify the accuracy of DFT+DMFT method, we first performed ion relaxation on the lattice structure of CeCoSi obtained experimentally, using the DFT, and DFT+DMFT (T = 300 K) methods, at different pressures. And calculated the system energy at different volumes under these three methods. The equilibrium volume was determined by fitting the *E-V* curve using the Birch-Murnaghan state equation. Figure 1(b) has shown the *E-V/V*$_0$ curve calculated by DFT+DMFT ($V_0$ is the volume at ambient pressure from the experiment), along with the equilibrium volumes obtained by the different methods. The equilibrium volume by DFT methods deviate significantly from the experimental values ($V_{\text{DFT}}/V_0$ = 0.925). However, the DFT+DMFT method can more accurately describe the correlation effects of Ce-4*f* electrons, and the obtained values were in good agreement with the experimental result ($V_{\text{DFT+DMFT}}/V_0$ = 0.997), proving the reliability of this method. In addition, the lattice constants of CeCoSi obtained by the two methods were significantly different, suggesting that the properties of Ce-4*f* electrons may be closely related to the crystal structure. Given that result of DFT+DMFT agrees well with the experimental equilibrium volume, we used the experimentally determined lattice constants for our subsequent electronic structure calculations.

Before exploring the effect of pressure on the electronic properties of CeCoSi using DFT+DMFT, we first studied the evolution of electronic structure with temperature at ambient pressure. Fig. 2 has shown the spectral function and density of states (DOS) of CeCoSi by DFT+DMFT at different temperatures under ambient pressure. The conduction band near $E_F$ is primarily composed of Co-3*d* electrons. At a high temperature of 500 K, very small Ce-4$f_{5/2}$ and Ce-4$f_{7/2}$ peaks of DOS appear at ~ 0.1 eV and ~ 0.4 eV, and hybridization between the Ce-4*f* and Co-3*d* bands is not evident in the spectral function. As temperature decreases, the Ce-4$f_{5/2}$ and Ce-4$f_{7/2}$ peaks sharpen, and hybridization becomes visible in the spectral function. To quantify the electron correlation effects of the Ce-4$f$ orbitals, we also calculated the renormalization factor $Z = m_{\text{DFT}}/m^*$ for Ce-4$f_{5/2}$ and Ce-4$f_{7/2}$ at different temperatures. $Z$ were calculated using the self-energy at Matsubara frequencies, where $Z^{-1} = 1 - \frac{\partial \text{Im}\Sigma(i\omega_n)}{\partial \omega_n}|_{\omega_n \to 0}$, thus avoiding the error caused by the maximum entropy method. Specifically, a three-order polynomial was fit to the lowest six Matsubara self-energy points and extrapolated to zero frequency. As temperature decreases, $Z$ of Ce-4$f_{7/2}$ changes slightly (0.40 at 500 K, and 0.44 at 100 K), while $Z$ of Ce-4$f_{5/2}$ decreases sharply (0.41 at 500 K, and 0.05 at 100 K). This suggests that, with decreasing temperature, the Kondo effect strengthens, leading to enhanced hybridization and an increase in the effective mass of the Ce-4$f_{5/2}$ quasiparticles. These results indicate that CeCoSi is in the Kondo metal state at ambient pressure, with increasing strength of hybridization as temperature decreases.

To study the changes in the electronic structure before and after the structural phase transition of CeCoSi, we performed DFT+DMFT calculations at various pressures. Figure 3 shows the evolution of electronic structure at low temperature (100 K) with pressure. As shown in Fig. 3(a), with increasing pressure, the hybridization between Ce-4$f$ and the conduction band is significantly enhanced. At 6 GPa, the distinct hybridized bands of Ce-4$f_{5/2}$ appears above $E_F$ about 0.05 eV. As pressure increases further, the hybridized bands widen and shift slightly upward. The same trend can be observed in the DOS shown in Fig. 3(b). As pressure increases, the Ce-4$f$ peak, especially Ce-4$f_{5/2}$ near the $E_F$, grows significantly, and its width also increases. Figure 3(c) has shown the evolution of $Z$ for Ce-4$f_{5/2}$ and Ce-4$f_{7/2}$ with pressure, at 100 K. As pressure increases, the hybridization strength of Ce-4$f_{5/2}$ increases, but the effective mass of the quasiparticle decreases. These results suggest that as the distance between Ce and Co ions decreases, the kinetic energy of Ce-4$f$ electrons increases, weakening the Kondo screening. We also calculated the Fermi surface of CeCoSi at different pressures using DFT+DMFT (see Fig. S1 in the Supplementary Materials[37]). Within the pressure range of 4 to 6 GPa, the Fermi surface undergoes a Lifshitz transition, at 100 K. These findings collectively indicate that CeCoSi experiences a pressure-induced quantum phase transition. Figure 4 has shown the valence state histogram given by CT-HYB at 100 K, which reflects the lifetime of atomic eigenstates. At ambient pressure, the atomic state with $|N=1, J=2.5\rangle$ was absolutely dominant, accounting for about 89.3%, at 100 K. As the pressure increases, the probability of other atomic states increases rapidly, reflecting the instability of the Ce valence state under pressure. It suggests that as pressure increases, CeCoSi transitions from the Kondo metal to a mixed-valence state.

To investigate the effect of temperature on the phase transition of CeCoSi, we also performed DFT+DMFT calculations at higher temperatures (300 K and 500 K) at various pressures. As shown in Fig. 5(a), at ambient pressure, $n_f$ is around 1.015 at different temperatures; $n_f$ decreases as pressure increases, but the rate of decrease was much more pronounced at lower temperatures. The change rate of $n_f$ with strain ($dn_f/d(V/V_0)$) first increases and then decreases with increasing pressure. It should be noted that the quantum phase transition from Kondo metal to Mixed-valence state is a gradual process with the Kondo screening gradually weakens. So, we used the pressure when $dn_f/d(V/V_0)$ reaches a maximum as transition pressure $P_Q$[10]. At 100 K, $P_Q$ is about 5.5 GPa, coinciding with the results of the Fermi surface (see supplementary materials[37]). As temperature increases, $P_Q$ shifts upwards, reaching approximately 6.5 GPa at 300 K. Figure 5(b) has shown the variation of the probabilities $P_{N=0}$, $P_{N=1}$ and $P_{N=2}$ in the Ce-4$f$ impurity model with pressure at different temperatures. At ambient pressure, $P_{N=0}$, $P_{N=1}$ and $P_{N=2}$ basically do not change with temperature. As pressure increases, $P_{N=0}$ increases and $P_{N=1}$ decreases, with more significant changes at lower temperatures. We also calculated the difference between DFT+DMFT and the superposition of spherical atomic electron density at different pressures and temperatures (see Fig. S2 in supplementary materials[37]). As pressure increases and temperature decreases, the electron density of Co and Si ions remains nearly constant, while Ce ions lose more electrons. These results suggest that the Ce ions gradually shift from $Ce^{3+}$ at low pressures to a mix state of $Ce^{3+}$ and $Ce^{4+}$ at high pressures, with high temperatures inhibiting this transition. As pressure increases, CeCoSi undergoes a quantum phase transition from Kondo metal to mixed-valence state, and the higher the temperature, the higher the pressure required for the transition.

We compared the quantum phase transition pressure $P_Q$ with the experimentally observed structural phase transition pressure $P_S$ at different temperatures, as shown in Fig. 5(c). The $P_Q$ was slightly higher than $P_S$ ($P_Q \sim 6.5$ GPa and $P_S \sim 4.9$ GPa at 300 K, respectively), and the temperature-dependent trends of both is the same. As pressure increases, CeCoSi undergoes a structural phase transition before entering the mixed-valence state, suggest that the valence instability of Ce-4$f$ is closely related to the structural phase transition. The increase in pressure causes the electron distribution of Ce-4$f$ orbitals to become more delocalized, leading to a redistribution of electron density and a structural phase transition.

Finally, to explore the mechanism of the quantum phase transition, we calculated the hybridization function $\Delta(\omega)$ of Ce-4$f$ at different temperatures and pressures. The imaginary part of hybridization function Im$\Delta(\omega)$ can be used to describe the strength of hybridization. As shown in Fig. 6, for the mixed-valence state and Kondo metal state, we use solid and dashed lines, respectively. At 100 K, as the pressure increases, the -Im$\Delta(\omega)$ for Ce-4$f_{5/2}$ and Ce-4$f_{7/2}$ both increase significantly with a sharp peak emerges, near 0 eV. For the high-pressure mixed-valence state (6 GPa and 8 GPa), these peaks were very pronounced; while for the low-pressure Kondo metal, no significant peaks are observed near 0 eV. At pressures above 4 GPa, reducing the temperature significantly enhances the strength of hybridization; while at lower pressures, the temperature effect is less significant. At high temperatures, the pressure required to transform to the mixed-valence state was higher, but regardless of pressure and temperature, the hybridization strength of mixed-valence state was much stronger than that of the Kondo metal state. In summary, CeCoSi is a typical Kondo metal at low pressures. As pressure increases, the distance between Ce-Co ions decreases, significantly enhancing the strength of hybridization between Ce-4$f$ and Co-3$d$, leading to the delocalization of Ce-4$f$ electrons. This results in the transition to a mixed-valence state, where the occupation number of Ce-4$f$ electrons decreases. The valence instability of may lead a redistribution of electron density, ultimately inducing a structural phase transition.

## IV. CONCLUSION

In summary, we systematically investigated the electronic structure of CeCoSi at different temperatures and pressures by DFT and DFT+DMFT methods. The equilibrium volume obtained by DFT+DMFT was in excellent agreement with experimental results ($V/V_0 \approx 0.997$), indicating that the crystal structure of CeCoSi is closely related to the strong correlation effects of Ce-4$f$ electrons. At ambient pressure, CeCoSi is a Kondo metal with hybridization between Ce-4$f$ and Co-3$d$ conduction bands can be observed in the spectral function. The occupation number of Ce-4$f$ is highly sensitive to pressure. As pressure increases, the occupancy of Ce-4$f$ decreases sharply, and CeCoSi transitions into mixed-valence state. At 100 K, the pressure of quantum phase transition $P_Q$ is about 5.5 GPa. As the temperature increases, $P_Q$ increases, reaching about 6.5 GPa at 300 K. The experiment results shown that CeCoSi undergoes a structural phase transition near 3.6 GPa at 10 K, with the $P_S$ increasing with temperature, reaching 4.9 GPa at 300 K. The $P_Q$ by our DFT+DMFT calculation was slightly higher than $P_S$, by experiment and both exhibit a same temperature dependence. These suggest that the pressure-induced valence instability of Ce-4$f$ drives the structural phase transition. As pressure increases, the distance between Ce-Co ions decreases, the strength of hybridization was significantly enhanced, and Ce-4$f$ electrons become more delocalized, leading CeCoSi to transform into the mixed-valence state. This valence instability may cause the redistribution of electron density and electron cloud, thereby inducing a structural phase transition. Our work is of significant importance for understanding the quantum and structural phase transitions in $R$CoSi compounds. However, due to the lack of experimental high-pressure structural measurements for CeCoSi, its lattice structure under high pressure remains unclear, and further studies on the electronic structure of its high-pressure phase are needed in future.

## ACKNOWLEDGMENTS

This work was supported by the National Natural Science Foundation of China (Grant No. 12474029, No. 12204033 and No. 12274117), the Program for Innovative

Research Team (in Science and Technology) in University of Henan Province (Grant No. 24IRTSTHN025), the Natural Science Foundation of Henan (Grant No. 242300421214), and the HPCC of HNU.

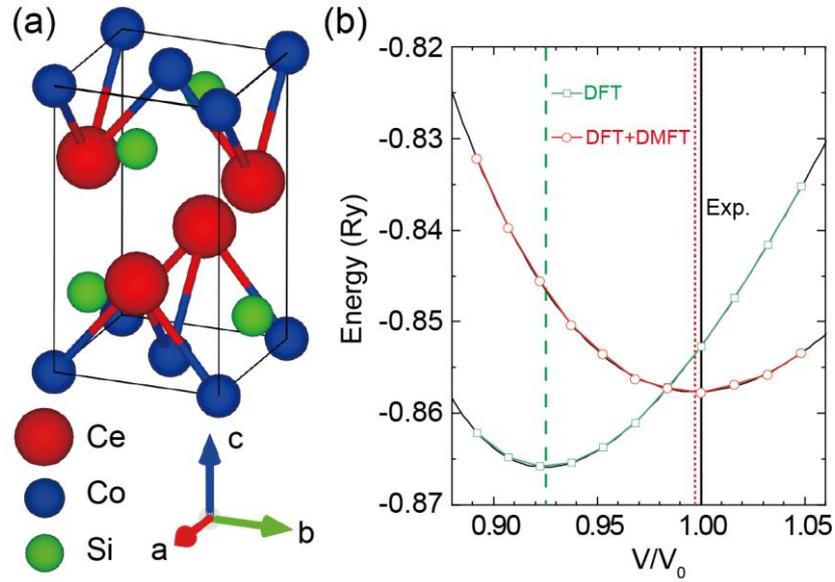

Fig. 1. (a) Crystal structure of CeCoSi. Red, blue and green balls represent Ce, Co, and Si atoms, respectively. (b) Calculated $E-V$ relation of CeCoSi by DFT and DFT+DMFT. A reference energy (−42195 Ry) was subtracted from the total-energy data. Here and in the following, $V/V_0$ means the volume compression ($V_0$ represents the experimental crystal volume[23]). The three vertical lines from left to right represent the equilibrium volumes by DFT, DFT+DMFT and experiment respectively.

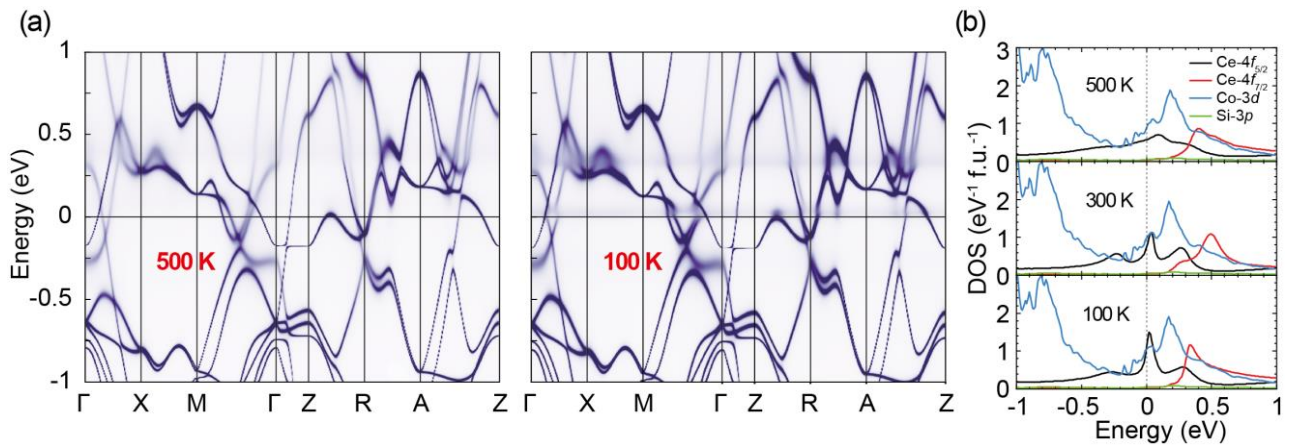

Fig. 2. The electronic structure of CeCoSi under ambient pressure by DFT + DMFT. (a) The momentum-resolved spectral functions, at 500 K and 100 K. (b) The spectral density at 500 K, 300 K, and 100 K. The black, red, blue and green solid lines represent Ce-$4f_{5/2}$, Ce-$4f_{7/2}$, Co-$3d$ and Si-$3p$ contributions, respectively.

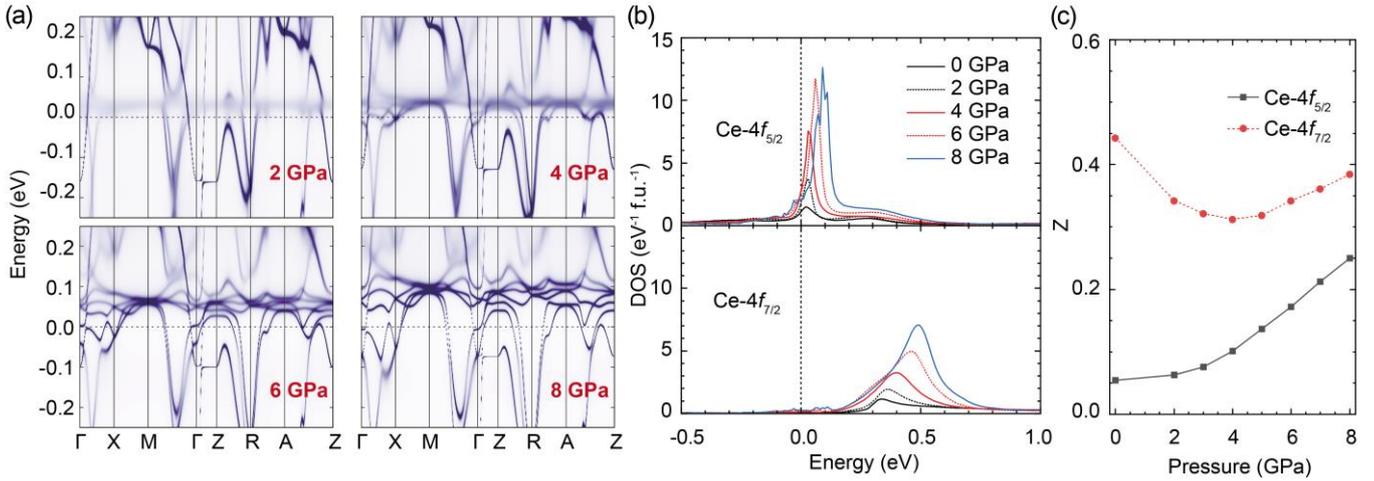

Fig. 3. The evolution of CeCoSi electronic structure with pressure at low temperature. (a) The momentum-resolved spectral functions and (b) spectral density under different pressures, at 100 K. The black solid lines, black dotted lines, red solid lines, red dotted lines and blue solid lines represent 0 GPa, 2 GPa. 4 GPa, 6 GPa and 8GPa, respectively. (c) The renormalization factor $Z$ of Ce-4$f_{5/2}$, Ce-4$f_{7/2}$ at different pressure.

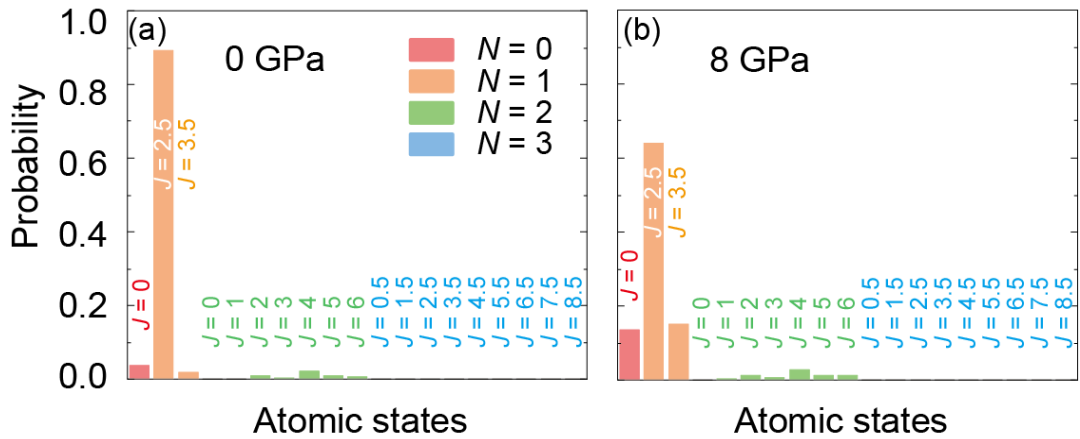

Fig. 4. Distributions of atomic state probability for Ce-4$f$ states at (a) 0 GPa and (b) 8 GPa obtained by DFT +DMFT calculations (100 K). The percentages for the $N = 3$ atomic states are too small (<1%).

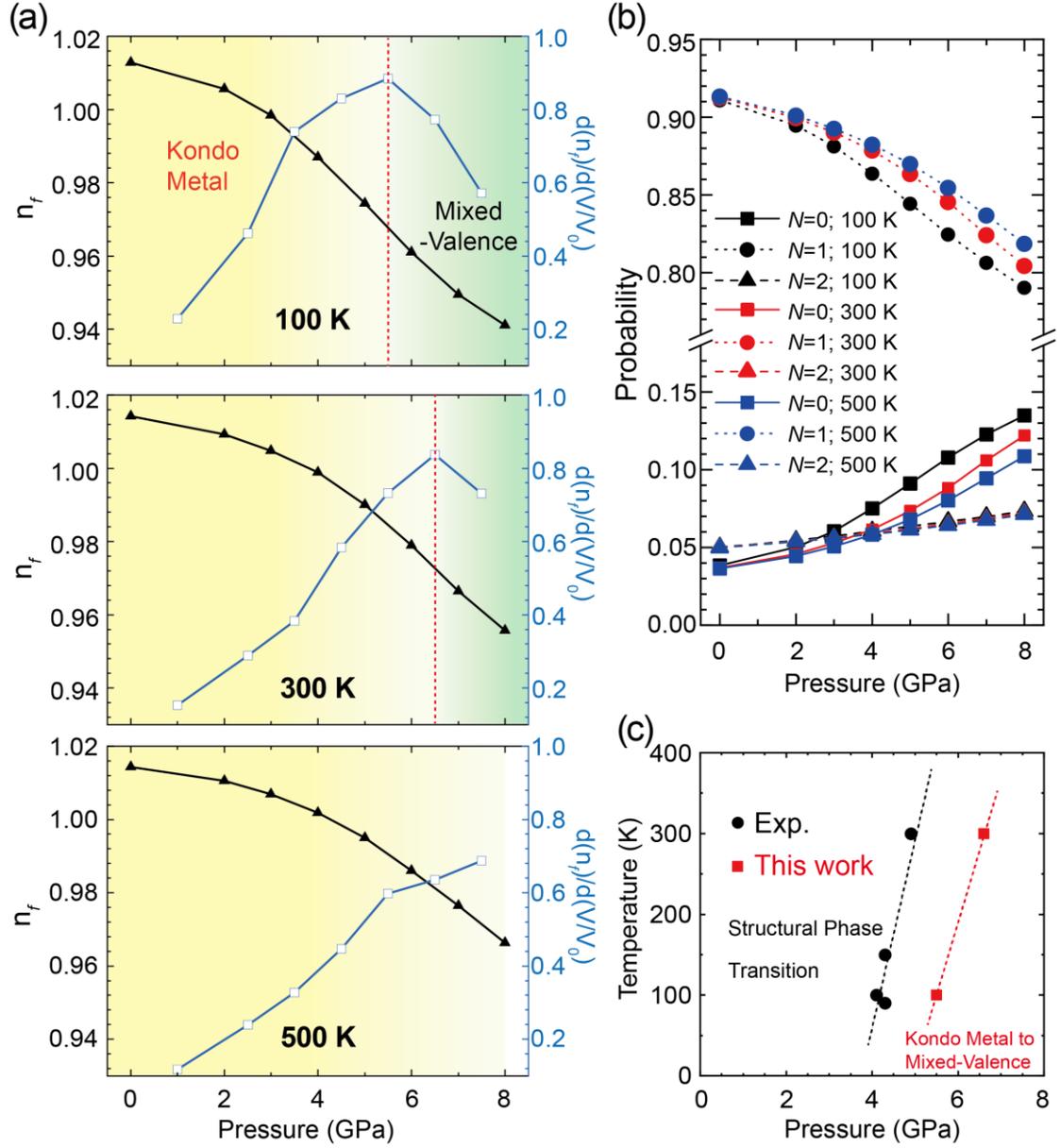

Fig. 5. Effect of temperature on phase transition pressure of CeCoSi by DFT+DMFT. (a) The number of occupied Ce-4$f$ electrons $n_f$ and derivative $d(n_f)/d(V/V_0)$ at 100 K, 300 K and 500 K. (b) The probability for Ce-4$f$ states at 100 K, 300 K and 500 K. (c) The structural phase transition by X-ray diffraction experiment[23] and quantum structure phase transition pressure by DFT+DMFT, at different temperature.

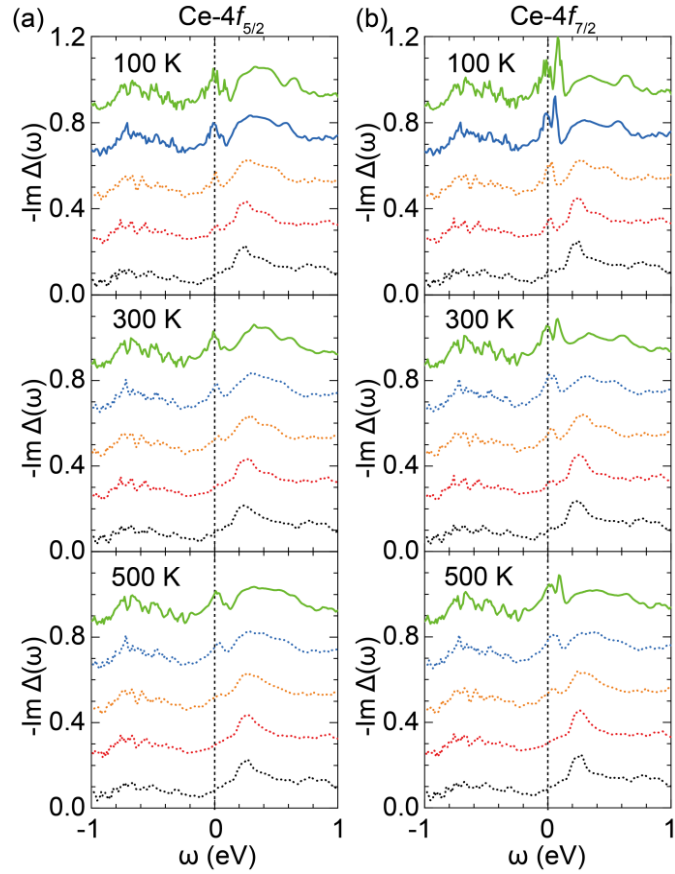

Fig. 6. The imaginary part of hybridization function on real frequency Im$\Delta(\omega)$ at different temperature and pressure, obtained by DFT + DMFT. (a) For the Ce-4$f_{5/2}$ states. (b) For the Ce-4$f_{7/2}$ states. The black, red, orange, blue and green lines represent 0, 2, 4, 6 and 8 GPa respectively (from bottom to top). The dashed lines and solid lines represent the Kondo metal state and mixed-valence state, respectively.